# X-ray interference effects on the determination of structural data in ultrathin La$_{2/3}$Sr$_{1/3}$MnO$_3$ epitaxial thin films

D. Pesquera, R. Bachelet, G. Herranz, J. Fontcuberta

*Institut de Ciencia de Materials de Barcelona, (ICMAB-CSIC), Campus de la UAB, 08193 Bellaterra, Spain*

X. Marti, V. Holy

*Department of Condensed Matter Physics, Faculty of Mathematics and Physics, Charles University, Ke Karlovu 5, 121 16 Prague 2, Czech Republic*

**Abstract**

*We analyze X-ray diffraction data used to extract cell parameters of ultrathin films on closely matching substrates. We focus on epitaxial La$_{2/3}$Sr$_{1/3}$MnO$_3$ films grown on (001) SrTiO$_3$ single crystalline substrates. It will be shown that, due to extremely high structural similarity of film and substrate, data analysis must explicitly consider the distinct phase of the diffracted waves by substrate and films to extract reliable unit cell parameters. The implications of this finding for the understanding of strain effects in ultrathin films and interfaces will be underlined.*

An intense research activity has been developed in the last years in the field of interfaces in oxide-based heterostructures due to the emerging properties that could promote. A celebrated example is the high conducting electron gas at the SrTiO$_3$/LaAlO$_3$ (STO/LAO) interfaces [1]. While the origin of this phenomenon has been attributed to the polar discontinuity, some other effects might be relevant. In particular, stoichiometry deviations at the interface have been observed [2, 3] which affect unit cell dimensions and thus the position of the Bragg reflections in X-ray diffraction experiments [4]. A second example are the ferromagnetic La$_{2/3}$A$_{1/3}$MnO$_3$ (A= Ca, Sr, Ba) manganites, where the competing coupling of spin, charge, orbital and lattice degrees of freedom may be largely affected by interface phenomena. Indeed, a *dead* layer has been identified in ultrathin La$_{2/3}$A$_{1/3}$MnO$_3$ films, causing the depression of the magnetic and conducting properties of the films [5]. The microscopic origin of this *dead* layer has been attributed to various sources, including orbital reconstruction [6,7] and cation disproportionation at the interfaces [8, 9, 10]. X-ray diffraction (XRD) has been used to show that ultrathin La$_{2/3}$A$_{1/3}$MnO$_3$ films display anomalous variations in cell volume [11], changes of structural symmetry [12], or modifications of MnO$_6$ octahedra rotation pattern [13].

Here we address the possible occurrence of structural modifications at interfaces in ultrathin manganite thin films by means of high resolution θ-2θ X-ray diffraction measurements. It will be shown that for an accurate description of the diffraction patterns, the superposition of the scattered waves by the substrate and film should be considered rather than a simple addition of intensities. Using this approach, accurate out-of-plane cell parameters of the films can be extracted; they largely differ from those obtained using conventional data analysis methods. The implications of these findings for restructuration at interfaces are discussed.

A set of La$_{2/3}$Sr$_{1/3}$MnO$_3$ (LSMO) films with thicknesses (*t*) ranging from 6 to 27 nm have been grown by pulsed laser deposition on (001)-oriented single crystalline SrTiO$_3$ (STO) substrates [7]. XRD measurements have been done by using a X'Pert diffractometer (Cu-Kα radiation). Reciprocal space maps (not shown here) indicate a fully coherent cube-on-cube epitaxial growth.

In Fig. 1(a-c) we show θ-2θ diffraction scans around the (002) reflection for films of 10 nm, 17 nm and 27 nm, respectively, showing the Bragg peaks ($\theta_B$) for substrate and layer and the Laue oscillations. Data analysis can be attempted by fitting the film contribution to the diffraction pattern to the expected intensity dependence of Laue's oscillations:

$$I(Q) = \frac{\sin^2\left(\tfrac{1}{2}QNc\right)}{\sin^2\left(\tfrac{1}{2}Qc\right)} \tag{1}$$

, where N is the number of unit cells along the [00$l$] direction, Q is the reciprocal lattice vector, given by $4\pi\sin(\omega)/\lambda$ where ω is the angle of the incident X-rays respect to the diffracting crystal planes. From the fit, both the thickness ($t$) – related to N – and $c$-axis parameter of the film – related to maximum of intensity at the Bragg angle, i.e. for Q(ω = $\theta_B$) – can be extracted. Independently, the substrate peak could be adjusted to a pseudo-Voigt function centered in the reflection peak of the substrate. The results of these fits are shown (red solid lines) in Figs. 1 (a-c). It is implicitly assumed here that the diffraction patterns of the LSMO layer and STO substrate can be simply added. One can observe that the shape of the Bragg peaks for substrate and film can be well adjusted but there is a clear shift of the fitted fringes when compared to the experimental ones. This shift is clearly asymmetric, being more pronounced for ω < $\theta_B$ than for ω > $\theta_B$. A similar discrepancy was observed in related materials and attributed to strain-gradients in the films [14].

The thickness dependence of the $c$-axis values extracted using this approach are shown (solid circles) in Fig. 2. We first note that all $c$-values are smaller than the corresponding bulk value (0.3873 nm [15]) as expected from the tensile strain imposed by the STO substrate. However, data in Fig. 2 suggest that for the thinnest films, a gradual expansion of the unit cell occurs, which is at odds with the expected elastic deformation of the unit cell under tensile strain. This observation is in agreement with earlier results in LSMO [10] and $La_{2/3}Ca_{1/3}MnO_3$ [11].

We will show in the following that these systematic discrepancies arise from the fact that the assumed incoherent superposition of diffracted waves for film and substrate is not adequate for epitaxial ultrathin films grown on closely matching substrates [16].

The electric field of the substrate-diffracted beam $E_S$ (relative to the incoming beam) must be obtained from the dynamical theory. For a symmetric reflection, the angular dependence of $E_S$ (ω) can be written as [17]:

$$E_S(\omega) = -\left(\beta_S \pm \sqrt{\beta_S^2 - 1}\right) \tag{2}$$

, where $\beta_s$ is,

$$\beta_S = \frac{1}{\sqrt{C_{lS} \cdot C_{-lS}}}\left(\sin(2\theta_B)(\omega - \theta_B) + C_{0S}\right) \tag{3}$$

and $C_{lS}$ and $C_{0S}$ are the $l$- and 0-term of the Fourier expansion of substrate's crystal polarizability [18]. C's are complex numbers and thus $E_s$ is complex. $\theta_B$ is the Bragg angle of the (00$l$) reflection. The sign in Eq. 2 should be chosen such as $|E_S| < 1$, which is the physically relevant solution. In Fig. 3a we show $|E_s(\omega)|^2$ calculated for the (002) reflection of the STO substrate which nicely reproduces the measured patterns (see Figs. 1(a-c)).

The electric field amplitude of the film-diffracted beam $E_L$ can be obtained by either the kinematical or the dynamical approximation. The simplest kinematical approximation is usually employed [19], where $E_L$ is given by [17]:

$$E_L(\omega) = \frac{C_{lL}}{2\beta_L} \exp\left[\left(-i\frac{2\pi}{\lambda}\frac{\beta_L}{\sin(\theta_B)}t\right)-1\right] \quad (4)$$

, where $\beta_L$ is:

$$\beta_L = \frac{1}{\sqrt{C_{lL}C_{-lL}}}\left[\sin(2\theta_B)(\omega-\theta_B)+C_{0S}+2\varepsilon\sin^2(\theta_B)\right] \quad (5)$$

, being $t$ the thickness of the layer, $\lambda$ the wavelength of the X-rays; $C_{lL}$ and $C_{0L}$ are the $l$- and 0-term of the Fourier expansion of film polarizability. $\varepsilon = (c_L-c_S)/c_S$ is the relative difference of out-of-plane cell parameters of substrate and film. In Fig. 3b we show $|E_L(\omega)|^2$, obtained using Eq. 4, for a LSMO layer of $t = 20$ nm, and $\varepsilon = -2.4$ % (appropriate for a fully strained LSMO on STO). It can be appreciated that $|E_L(\omega)|^2$ displays the characteristic Laue oscillations. Indeed, it can be shown that $|E_L(\omega)|^2$ is identical to I(Q) given by Eq. 1.

In the case of interest here, i.e, a coherent film on a single crystalline substrate, the electric field amplitude of the diffracted beam by the substrate-film $E_{SL}$ must be obtained by computing $|E_{SL}(\omega)|^2 = |E_S(\omega) + E_L(\omega)|^2$ with $E_S(\omega)$ and $E_L(\omega)$ as given by Eqs. 2 and 4 [20]. The interference of the $E_S(\omega)$ and $E_L(\omega)$ diffracted beams strongly modifies the computed patterns and produces a shift of the measured maxima and an asymmetric intensity patterns at $\omega > \theta_B$ and $\omega < \theta_B$, as indicated by red-line in Fig. 3c. $|E_{SL}(\omega)|^2$ differs significantly from the pattern calculated assuming no-interference of the diffracted beams ($|E^0_{SL}(\omega)|^2 = |E_S(\omega)|^2 + |E_L(\omega)|^2$) shown by blue-lines in Fig. 3c.

In Fig. 1d-f (dashed lines), we show the results of the corresponding fits obtained using $|E_{SL}(\omega)|^2 = |E_S(\omega) + E_L(\omega)|^2$ as described above [21]. It can be appreciated that the fits are excellent and all asymmetries (maxima position and intensity) are well reproduced. The $c$-axes parameters extracted from these fits are included in Fig. 2 (open symbols). It is clear that when considering X-ray beam interference to fit the data, the extracted $c$-parameters display a monotonic increase when increasing thickness, as expected from a gradual tensile-strain relaxation.

In summary, we have derived simple expressions that can be used for accurate fitting of X-ray patterns of thin films. Its use, in the case of LSMO/STO heterostructures, allows extracting data that are at odds with previous reports showing a structural modification at the interfaces in manganites [11] and prove that the cell parameters of ultrathin films display a smooth and monotonic behavior as expected from substrate-induced elastic deformation of the lattice. Therefore, XRD do not support a change of chemical composition close to interfaces. Beyond the particular case of manganites analyzed here, the present results illustrate that the interaction of diffracted beams by substrate and films should be considered to extract reliable structural information in ultrathin films. To what extent these findings will affect claims of composition or structural at oxide interfaces can not be anticipated.

### Acknowledgements

Financial support by the Ministerio de Ciencia e Innovación of the Spanish Government [Projects MAT2008-06761-C03, MAT2011-29269-CO3 and NANOSELECT CSD2007-00041] and Generalitat de Catalunya (2009 SGR 00376) is acknowledged.

**Figure Captions**

Fig 1. (Color online) θ-2θ X-ray diffraction patterns of selected LSMO/STO samples. Points are experimental data. In (a-c) solid lines correspond to results of fits by adding the film and substrate contributions (dashed lines); In (d-f) solid lines are fits using the square of the sum of the electric fields amplitudes $|E_{SL}(\omega)|^2 = |E_S(\omega) + E_L(\omega)|^2$

Fig 2. Out-of-plane lattice parameters for LSMO films, calculated either by means of (black circles) Bragg's law from the position of the diffraction peak of the layer or (open circles) by fitting the data using the square of the sum of the electric fields amplitudes $|E_{SL}(\omega)|^2 = |E_S(\omega) + E_L(\omega)|^2$. Dashed line indicates the bulk value of *c*-axis of LSMO.

Fig 3. (Color online) (a, b) Amplitudes of the diffracted waves by the STO substrate (Eq. 2) and a LSMO layer (Eq. 4) respectively. c) Total diffracted intensity calculated as (blue) $|E^0_{SL}(\omega)|^2 = |E_S(\omega)|^2 + |E_L(\omega)|^2$ and (red) $|E_{SL}(\omega)|^2 = |E_S(\omega) + E_L(\omega)|^2$

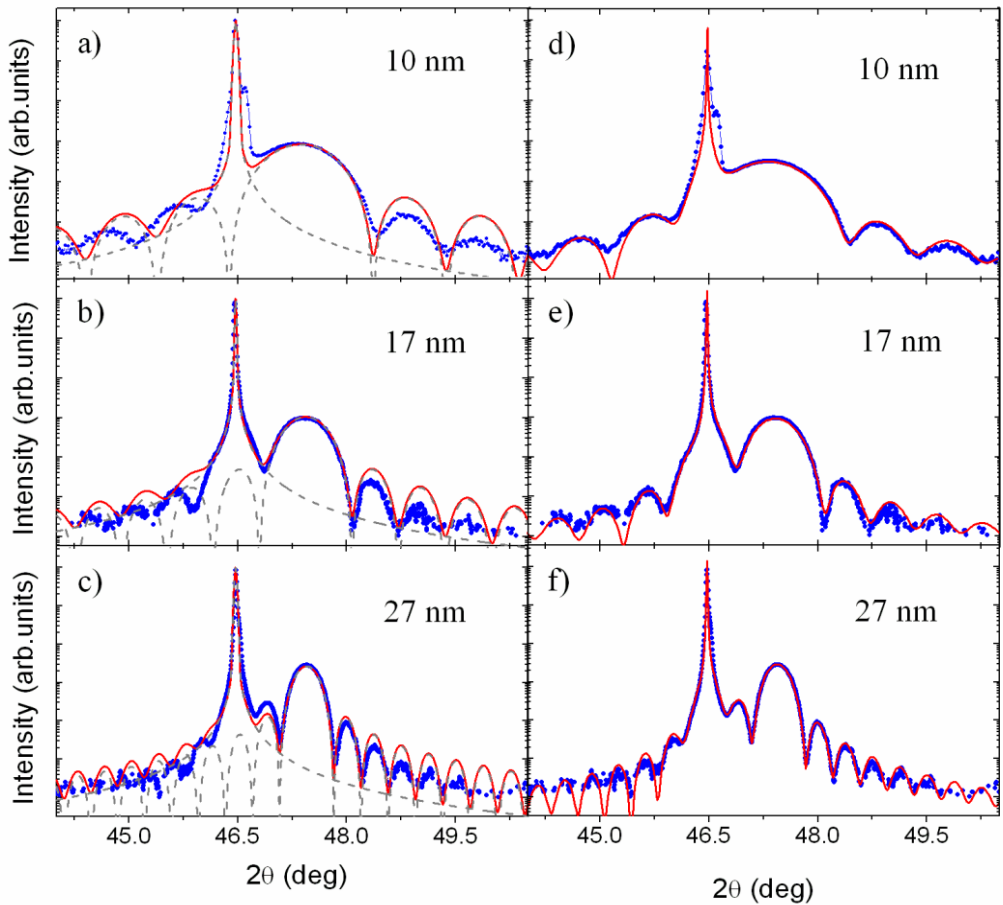

Fig. 1

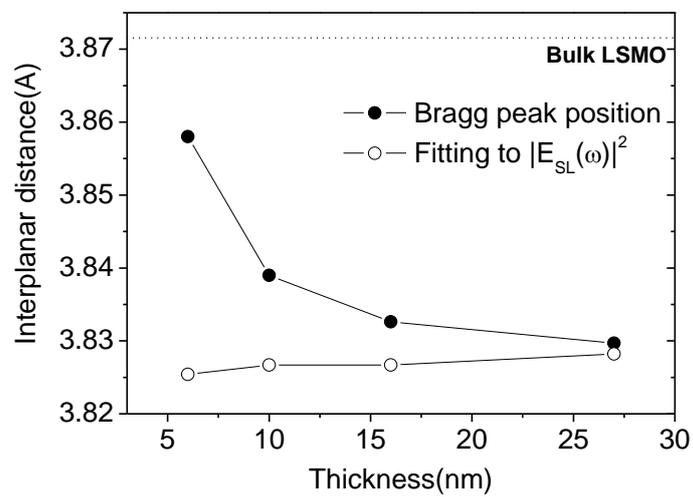

Fig. 2

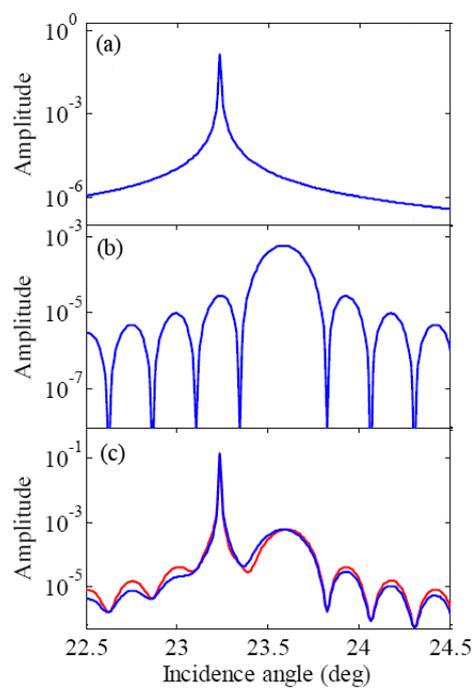

Fig. 3